# GNU-Octave Como Alternativa de Simulación de Sistemas Dinámicos No Lineales en la Enseñanza de la Ingeniería.

TORRES Felipe de Jesús†*, ARREDONDO Monserrat Sugey, MARTINEZ José Manuel, OCAMPO Víctor Manuel.

*Universidad de Guanajuato, División de Ingenierías Campus Irapuato-Salamanca.*

___

**Resumen**

Este artículo presenta una propuesta de alternativa para simular sistemas dinámicos no lineales, en la enseñanza de la ingeniería. Teniendo aplicación en programas de licenciatura como: Ingeniería Electromecánica, Ingeniería en Redes y Telecomunicaciones, Ingeniería Mecánica, entre otras que se imparten en las distintas universidades públicas del estado de Guerrero. Los equipos de cómputo, comúnmente utilizados para las simulaciones, requieren de gran capacidad de hardware para soportar el software de simulación. Aunado a esto, el software de simulación en la mayoría de los casos, son bajo licencia. Por tanto, el costo de implementar un laboratorio de simulación en una universidad pública es muy alto. Así, se muestra una alternativa de usar una tarjeta de desarrollo comercial *Raspberry Pi*, con el software *GNU-Octave* de licencia libre, para simular sistemas dinámicos no lineales como un robot *SCARA* de 4 grados de libertad y un péndulo invertido rotacional. La comparación de los modelos dinámicos simulados tanto en software especializado como en el software libre propuesto, exhiben la viabilidad de la alternativa propuesta.

**Palabras clave:** simulación, sistemas dinámicos, no lineal, Raspberry Pi, GNU-Octave.

**Abstract**

This paper presents a proposed alternative to simulate non-linear dynamical systems. This has an application in bachelor programs like: Electrical and mechanical engineering, Networks and Telecommunications engineering, Mechanical engineering and more, than they are taught in several public universities in Guerrero state. Commonly, the computer devices used for simulations require of high hardware capacity to support the simulation software. Moreover, the simulation software in the majority of the cases is under licence permission. For these reasons, implementing a simulation lab in a public university is very high cost. Thus, we show an alternative by using a commercial development board *Raspberry Pi* supporting the *GNU-Octave* software, which is a free software, to simulate non-linear dynamical systems like a 4 grades of freedom *SCARA* robot and a rotational inverted pendulum. The comparision of the simulated dynamical models in both the specialized software and the proposed free software, exhibit the viability of the proposed alternative.

**Keywords:** simulation, dynamical systems, non-linear, Raspberry Pi, GNU-Octave.

___

___

*Correspondencia al Autor (fdj.torres@ugto.mx)

† Investigador contribuyendo como primer autor.

## I.     Introducción

Las estrategias de enseñanza-aprendizaje en la educación superior actual, están relacionadas cada vez más con el uso de las Tecnologías de Información y Comunicación (TIC). Particularmente, en la enseñanza de la Ingeniería es necesario utilizar equipos de cómputo para realizar simulaciones de diversa índole. Por ejemplo, en cursos como Sistemas de Control, Sistemas Dinámicos, Robótica, Ecuaciones Diferenciales, Matemáticas Avanzadas, entre otras, se llevan a cabo simulaciones de sistemas dinámicos lineales y no lineales.

En varias de las universidades públicas del estado de Guerrero se imparten licenciaturas en Ingeniería Eléctrica y Mecánica, Redes y Telecomunicaciones, Energía, Sistemas Computacionales, Civil, entre otras que requieren de la puesta en marcha de un laboratorio de simulación para que el estudiante pueda realizar las simulaciones de los sistemas dinámicos no lineales tratados en el aula. Para el funcionamiento adecuado del laboratorio de simulación, el equipo de cómputo a utilizar debe cumplir, generalmente, con gran capacidad de hardware para soportar los softwares de simulación especializados. Estas computadoras por tanto son de tamaño grande y costosas. Otra problemática que afrontan los administrativos de las escuelas públicas es la compra de licencia de los softwares de simulación, las cuales son igualmente costosas y están en función de la cantidad de equipos de cómputo donde serán instaladas, además de tener una vigencia finita, lo cual representa un gasto periódico.

Existen otro tipo de software denominado software libre, el cual es un movimiento que se ha gestado desde principios de los 70's, derivado del posicionamiento de las grandes empresas de empezar a vender los softwares que producían. Aunque fue hasta principios de los 80's cuando el movimiento de software libre inició a producir sus propios softwares, los cuales estaban bajo una licencia pública denominada GPL (General Public License, por sus siglas en inglés). En Viñas (2003) expresa que ésta licencia protegía al software desarrollado a partir del movimiento de software libre, otorgándole las libertades de:

- Usar el software para cualquier propósito.
- Estudiar cómo funciona el software y adaptarlo a las necesidades propias.
- Distribuir libremente copias del software.
- Mejorar el software y hacer públicas las versiones mejoradas en beneficio de la comunidad.

Las ventajas de utilizar software libre están basadas en el costo de la licencia que se elimina completamente, respecto a un software de simulación especializado. Más aún, si éste software libre es instalado en una minicomputadora o bien, una tarjeta de desarrollo comercial como la *Raspberry Pi,* la cual tiene un costo 20 veces menor al costo de una computadora personal comúnmente usada para simulación.

Sin embargo, la problemática de utilizar software libre para simulación de sistemas dinámicos no lineales se centra en que la información disponible en la literatura para el proceso de enseñanza-aprendizaje, está en función del uso de software especializado como Matlab$^{TM}$. Así por ejemplo en (Houpis, 2013) se presenta una guía de simulación de sistemas dinámicos lineales con Matlab$^{TM}$. En (Khatib, 2016) se modelan sistemas fotovoltaicos usando Matlab$^{TM}$. En (Blaabjerg, 2017) se simulan sistemas de energías renovables a través de Matlab$^{TM}$. En (López, 2014) trata acerca de aplicaciones de la ingeniería en sistemas de control con Matlab$^{TM}$.

Por todo lo anterior, es posible mostrar que un software libre, soportado en una tarjeta de desarrollo como la Raspberry Pi, puede ser usado como software de simulación de sistemas dinámicos no lineales.

El artículo se estructura de la siguiente manera: en la sección II se presenta la descripción del software GNU-Octave y de la Raspberry Pi. En la sección III se muestran los modelos dinámicos no lineales del robot SCARA y del péndulo invertido rotacional. En la sección IV, se exhiben los resultados de las simulaciones realizadas en el software comúnmente usado, Matlab$^{TM}$ y el software libre propuesto. Por último, en la sección V se dan las conclusiones del documento.

**Objetivos**

- Simular sistemas dinámicos no lineales en software libre soportado en una minicomputadora o tarjeta de desarrollo como Raspberry Pi.
- Evidenciar como alternativa de simulación en la enseñanza de la Ingeniería, el uso de GNU-Octave a través de la comparación con las simulaciones hechas en Matlab$^{TM}$.

## II. Descripción de GNU-Octave y Raspberry Pi.

### a. GNU-Octave.

GNU-Octave es un software libre redistribuible, lo cual permite ser modificado bajo los términos del GNU General Public License (GPL). Es un lenguaje de alto nivel para cálculo numérico. Permite la solución numérica de problemas lineales y no lineales. Provee extensas capacidades gráficas para visualización y manipulación de datos. Se usa a través de su línea de comandos interactivos o bien de códigos. El lenguaje Octave es bastante similar a Matlab$^{TM}$ de tal manera que la mayoría de los programas son portables.

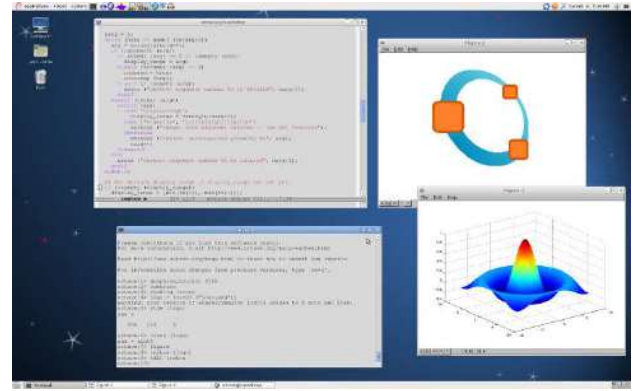

*Figura 1. Pantalla de GNU Octave.* Fuente: tomada de la pantalla principal del software GNU-Octave.

### b. Raspberry Pi

Raspberry PI® (RPi) es una mini-computadora en placa, de tamaño de una tarjeta de crédito, desarrollada en Reino Unido por la fundación Raspberry Pi con la intención de estimular la enseñanza de las ciencias computacionales básicas en escuelas (Sarthak, 2014). La fundación provee Debian y Python como lenguajes de programación principal, sin embargo, es posible instalar sistemas operativos en base a Linux o incluso Windows® 10. El bajo costo y la configuración de hardware de la placa RPi han hecho que sea muy popular entre los programadores y realizadores (aficionados) de proyectos de automatización que requieren algún procesamiento computacional. Está integrada por un chip Broadcom BCM2835 con procesador ARM hasta 1.4 GHz de velocidad, GPU VideoCore IV y hasta 1 GB de memoria RAM.

Se requiere de un medio de almacenamiento, tarjetas de memoria SD o microSD, así como de un cargador microUSB de al menos 2000mAh. Contiene un puerto de salida de video HDMI y otro de tipo RCA, minijack de audio y un puerto USB 2.0 (modelos A y A+, B dispone de dos USB y B+ y Raspberry Pi 2 disponen de 4 USB) con el que se podrá conectar periféricos como teclado y ratón. Para conexión en red, la RPi contiene un puerto Ethernet o es posible utilizar un adaptador inalámbrico WiFi compatible.

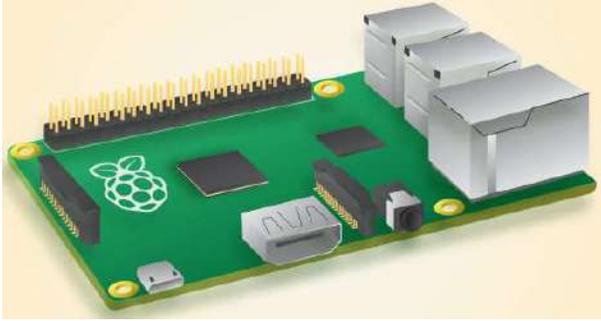

*Figura 2. Raspberry Pi B+.* Fuente: https://www.raspberrypi.org/wp-content/uploads/2013/12/model-b-plus.png.

### III. Modelos Dinámicos de Sistemas No-lineales.

#### a. Modelo Dinámico del Robot SCARA

El robot SCARA (Selective Compliance Articulated Robot Arm, por sus siglas en inglés) es un brazo robótico de configuración revoluta (rotacional o de pasador) horizontal en sus articulaciones $q_1$, $q_2$ y $q_3$, como se observa en la Figura 3. Fue diseñado en Japón y desempeña tareas de tipo "tomar y poner", por lo que es útil en líneas de ensamble.

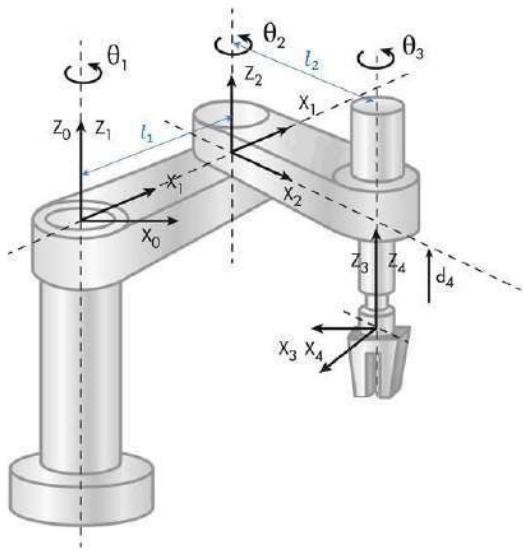

*Figura 3. Robot manipulador SCARA.* Fuente: Elaboración propia.

Se considera un robot manipulador rígido completamente actuado, despreciando las pérdidas por fricción, con $i = 1,2 \dots n$ articulaciones, donde $n = 4$. Los vectores de coordenadas generalizadas de las articulaciones del robot son $q_i \in \mathbb{R}^n$, bajo las siguientes características:

Los eslabones 1, 2 y 3 tienen un movimiento rotacional. Sus posiciones angulares $\theta_1$, $\theta_2$ y $\theta_3$ serán las coordenadas generalizadas $q_1$, $q_2$ y $q_3$, respectivamente. El desplazamiento vertical $d_4$ del efector final, será la coordenada generalizada $q_4$. Los movimientos $q_1$, $q_2$ y $q_3$ se desarrollan en el plano horizontal, por lo cual se desprecia la energía potencial $U(q_i)$ en ellos. La articulación 3 mueve las masas $m_3$ y $m_4$, así mismo se suman las inercias $I_3$ e $I_4$.

Usando el formulismo de Euler-Lagrange, el modelo dinámico del robot, en su representación matricial, es dado por:

$$M_i(q_i)\ddot{q}_i + C_i(q_i,\dot{q}_i)\dot{q}_i + B_i\dot{q}_i + g_i(q_i) = \tau_i \quad (1)$$

Donde $M_i(q_i) \in \mathbb{R}^{n \times n}$ es la matriz de inercias, $C_i(q_i,\dot{q}_i) \in \mathbb{R}^{n \times n}$ es la matriz de Coriolis y fuerzas centrífugas, $B_i$ es una matriz diagonal de los coeficientes de fricción viscosa de cada articulación, $g_i(q_i) \in \mathbb{R}^n$ es el vector de fuerzas gravitacionales, $\tau_i \in \mathbb{R}^n$ es el vector de torques o pares de entrada. De manera detallada,

$$\begin{bmatrix} M_{11} & M_{12} & M_{13} & M_{14} \\ M_{21} & M_{22} & M_{23} & M_{24} \\ M_{31} & M_{32} & M_{33} & M_{34} \\ M_{41} & M_{42} & M_{43} & M_{44} \end{bmatrix} \begin{bmatrix} \ddot{q}_1 \\ \ddot{q}_2 \\ \ddot{q}_3 \\ \ddot{q}_4 \end{bmatrix} + \begin{bmatrix} C_{11} & C_{12} & C_{13} & C_{14} \\ C_{21} & C_{22} & C_{23} & C_{24} \\ C_{31} & C_{32} & C_{33} & C_{34} \\ C_{41} & C_{42} & C_{43} & C_{44} \end{bmatrix} \begin{bmatrix} \dot{q}_1 \\ \dot{q}_1 \\ \dot{q}_3 \\ \dot{q}_4 \end{bmatrix} + \begin{bmatrix} B_1 & 0 & 0 & 0 \\ 0 & B_2 & 0 & 0 \\ 0 & 0 & B_3 & 0 \\ 0 & 0 & 0 & B_4 \end{bmatrix} \begin{bmatrix} \dot{q}_1 \\ \dot{q}_1 \\ \dot{q}_3 \\ \dot{q}_4 \end{bmatrix} + \begin{bmatrix} 0 \\ 0 \\ 0 \\ m_4 g \end{bmatrix} = \begin{bmatrix} \tau_1 \\ \tau_1 \\ \tau_3 \\ \tau_4 \end{bmatrix} \quad (2)$$

donde:

$$M_{11} = I_1 + I_2 + I_3 + I_4 + m_1 lc_1^2 + m_2 l_1^2 \\ + m_2(lc_2^2 + 2l_1 lc_2 cosq_2) \\ + (m_3 + m_4)(l_1^2 + l_2^2 \\ + 2l_1 l_2 cosq_2)$$

$$M_{12} = I_2 + I_3 + I_4 + m_2(lc_2^2 + l_1 lc_2 cosq_2) \\ + (m_3 + m_4)(l_2^2 + l_1 l_2 cosq_2)$$

$$M_{13} = M_{23} = M_{33} = I_3 + I_4$$

$$M_{14} = M_{24} = M_{34} = 0$$

$$M_{22} = I_2 + I_3 + I_4 + m_2 lc_2^2 + (m_3 + m_4)l_2^2$$

$$M_{44} = m_4$$

$$C_{11} = -m_2 l_1 lc_2 sinq_2 \dot{q}_2 \\ - (m_3 + m_4) l_1 l_2 sinq_2 \dot{q}_2$$

$$C_{12} = -m_2 l_1 lc_2 sinq_2 (\dot{q}_1 + \dot{q}_2) \\ - (m_3 + m_4) l_1 l_2 sinq_2 (\dot{q}_1 + \dot{q}_2)$$

$$C_{21} = m_2 l_1 lc_2 sinq_2 \dot{q}_1 \\ + (m_3 + m_4) l_1 l_2 sinq_2 \dot{q}_1$$

$$C_{13} = C_{14} = C_{22} = C_{23} = C_{24} = C_{31} = C_{32} \\ = C_{33} = C_{34} = C_{41} = C_{42} = C_{43} \\ = C_{44} = 0$$

Este modelo obtenido fue comparado con los modelos dinámicos presentados en (Voglewede, 2009) y (Lewis, 2006), los cuales son similares en función del orden de las coordenadas generalizadas.

### b. Modelo Dinámico del Péndulo Invertido Rotacional.

En base a (Duart, 2017), el modelo dinámico del péndulo invertido rotacional, mostrado en la Figura 4, se obtiene mediante las ecuaciones de Euler – Lagrange.

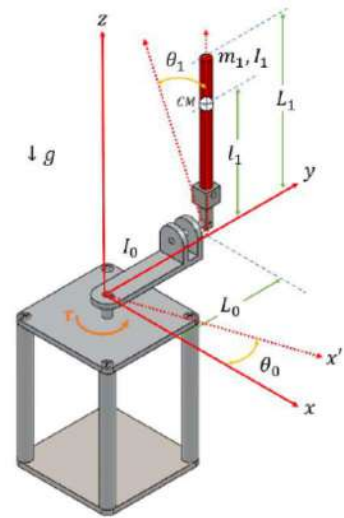

**Figura 4. Péndulo invertido rotacional.** *Fuente: Duart, 2017.*

Debido a que el sistema tiene dos grados de libertad éstas tienen la siguiente forma:

$$\frac{d}{dt}\left(\frac{\partial L}{\partial \dot{\theta}_0}\right) - \frac{\partial L}{\partial \theta_0} = \tau \quad (3)$$

$$\frac{d}{dt}\left(\frac{\partial L}{\partial \dot{\theta}_1}\right) - \frac{\partial L}{\partial \theta_1} = 0 \quad (4)$$

Donde $\tau$ es el par del motor y $L$ es el lagrangiano.

Este último, está definido como: $L = K_T - U_T$, el cual se refiere a la diferencia de la energía cinética total y la energía potencial total del sistema.

Por lo que, como primera instancia se requiere realizar el cálculo de la energía total tanto cinética como potencial del sistema. La energía cinética consiste de la componente traslacional y rotacional del péndulo y de la componente rotacional del brazo. Por otro lado, el sistema solo almacena energía potencial en el péndulo ya que el brazo, al tener movimiento rotacional en el plano horizontal, no presenta un cambio de altura en su centro de masa. Las ecuaciones resultantes son:

$$K_T = K_0 + K_1 = \frac{1}{2}I_0\dot{\theta}_0^{\,2} + \frac{1}{2}I_1\dot{\theta}_1^{\,2} \quad (5)$$
$$+ \frac{1}{2}m_1\Big(L_0^{\,2}\dot{\theta}_0^{\,2}$$
$$+ l_1^{\,2}\big(\dot{\theta}_1^{\,2} + \dot{\theta}_0^{\,2}\sin^2\theta_1\big)$$
$$+ 2L_0 l_1 m_1 \dot{\theta}_0 \dot{\theta}_1 \cos\theta_1\Big)$$

$$U_T = m_1 g l_1 (\cos\theta_1 - 1) \quad (6)$$

Por lo tanto, el lagrangiano es dado por:

$$L = \frac{1}{2}I_0\dot{\theta}_0^{\,2} + \frac{1}{2}I_1\dot{\theta}_1^{\,2} + \frac{1}{2}\big(l_0^{\,2} m_1 \dot{\theta}_0^{\,2}\big) \quad (7)$$
$$+ \frac{1}{2}\big(l_1^{\,2} m_1 \dot{\theta}_1^{\,2}\big)$$
$$+ \frac{1}{2}\big(l_1^{\,2} m_1 \dot{\theta}_1^{\,2} \sin^2\theta_1\big)$$
$$+ L_0 l_1 m_1 \dot{\theta}_0 \dot{\theta}_1 \cos\theta_1$$
$$+ m_1 g l_1 (1 - \cos\theta_1)$$

De las ecuaciones (3) y (4) se obtienen las ecuaciones de movimiento:

$$I_0 \ddot{\theta}_0 + L_0^{\,2} m_1 \ddot{\theta}_0 \quad (8)$$
$$+ \{l_1^{\,2} m_1 (\ddot{\theta}_0 \sin\theta_1$$
$$+ 2\dot{\theta}_0 \dot{\theta}_1 \sin\theta_1 \cos\theta_1)\}$$
$$+ \{L_0 l_1 m_1 \big(\ddot{\theta}_1 \cos\theta_1$$
$$- \dot{\theta}_1^{\,2} \sin\theta_1\big)\} = \tau$$

$$I_1 \ddot{\theta}_1 + l_1^{\,2} m_1 \ddot{\theta}_1 + L_0 l_1 m_1 \ddot{\theta}_0 \cos\theta_1 \quad (9)$$
$$- l_1^{\,2} m_1 \dot{\theta}_0^{\,2} \sin\theta_1 \cos\theta_1$$
$$- m_1 g l_1 \sin\theta_1 = 0$$

Así, (8) es la ecuación del movimiento del brazo y (9) del péndulo.

### IV. Comparación de Simulación en Matlab™ y GNU-Octave.

De manera general, las simulaciones son llevadas a cabo como un sistema de lazo abierto, donde el usuario introduce los pares de entrada que desee, éstos pares actúan sobre el modelo dinámico no lineal del robot SCARA o del Péndulo Invertido Rotacional. El software de simulación, sea Matlab™ o GNU-Octave resuelve el sistema de ecuaciones diferenciales a través de una doble integración para obtener la posición angular de la articulación. El esquema de simulación es visto en la Figura 5.

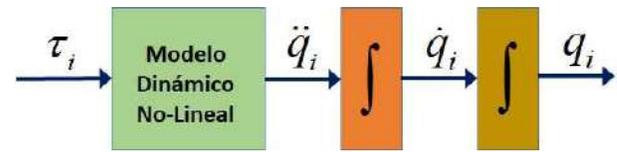

*Figura 5. Esquema general de simulación. Fuente: Elaboración propia.*

#### a. Simulación del Robot SCARA.

Se considera un robot manipulador de 4 grados de libertad tipo SCARA, utilizando el modelo dinámico (2). Los parámetros del robot SCARA don dados en la Tabla 1.

| Parámetro | Valor | Parámetro | Valor |
|---|---|---|---|
| $m_1$ | $15\,kg$ | $l_1$ | $0.50\,m$ |
| $m_2$ | $12\,kg$ | $l_2$ | $0.40\,m$ |
| $m_3 = m_4$ | $3\,kg$ | $l_{c1}$ | $0.25\,m$ |
| $I_1$ | $0.02 m_1$ | $l_{c2}$ | $0.20\,m$ |
| $I_2$ | $0.08 m_2$ | $I_3$ | $0.05 m_3$ |
| $I_4$ | $0.02 m_4$ | $g$ | $9.81\,\frac{m}{s^2}$ |
| $B_i$ | $0.5\,\frac{N\cdot s}{m}$ | | |

*Tabla 1: Parámetros del robot SCARA. (Fuente: Elaboración propia).*

### 1. Simulación en Matlab[TM]

Para realizar la simulación en Matlab[TM] se utilizó de la plataforma *Simulink®*, codificación de la *S-Function*, solucionador ode45 paso variable, con un tiempo de simulación de 10 segundos y condiciones inciales a cero. El esquema en *Simulink[TM]*, mostrado en la Figura 6, es sencillo en apariencia, lo importante es el código de la *S-Function*, el cual se detalla en el *anexo a*. Dentro del esquema de simulación se ha incluido un bloque $q\_dq$ que guarda el contenido de las variables de cada integración, en una estructura de arreglo con tiempo. Esta variable es usada para crear las gráficas correspondientes.

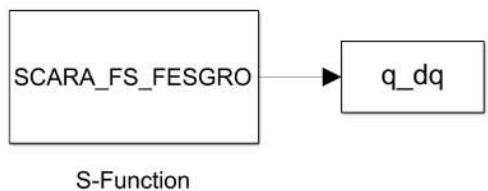

**Figura 6. Esquema general de simulación.** Fuente: Elaboración propia.

Así, en el Gráfico 1 se presenta el desplazamiento y velocidad angular para cada articulación del robot SCARA sometido a los pares siguientes:

$$\tau_i = \begin{bmatrix} 3 \\ 2 \\ 0 \\ 30 \end{bmatrix} N \cdot m \qquad (10)$$

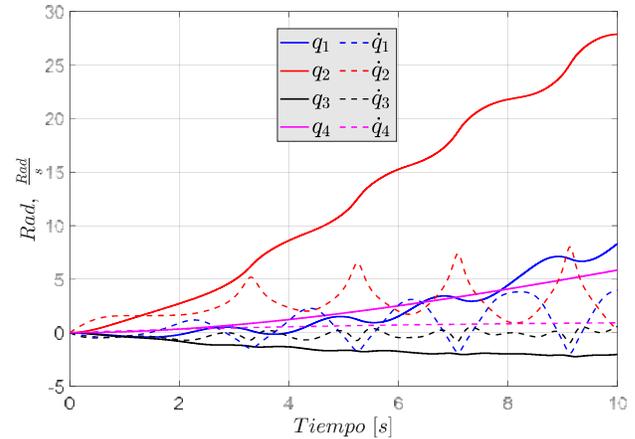

**Gráfico 1. Simulación en Matlab[TM] del sistema no lineal: robot SCARA.** Fuente: Elaboración propia.

### 2. Simulación en GNU-Octave

El procedimiento para llevar a cabo la simulación en GNU-Octave ha sido desarrollado de la siguiente manera:

1. Codificar una función que se ha denominado *projectile*, conteniendo:
   a. Los parámetros del robot.
   b. Los pares de entrada al sistema.
   c. Las ecuaciones diferenciales matriciales del modelo (2).
   d. Formar el vector de segundas y primeras derivadas que serán resueltas (integradas) por el solucionador elegido.
2. Crear un código que llame a la función *projectile*, éste código establece:
   a. Condiciones iniciales.
   b. Tiempo de simulación.
   c. Elección del solucionador.
   d. Propiedades del solucionador.
   e. Líneas de código para la creación de la gráfica correspondiente.

Estos códigos pueden ser analizados en el *anexo b*. Para realizar adecuadamente la comparación, en GNU-Octave se eligió el solucionador ode45 y un tiempo de simulación de 10 segundos, al igual que la simulación realizada en Matlab[TM]. La gráfica que GNU-Octave arroja es vista en el Gráfico 2.

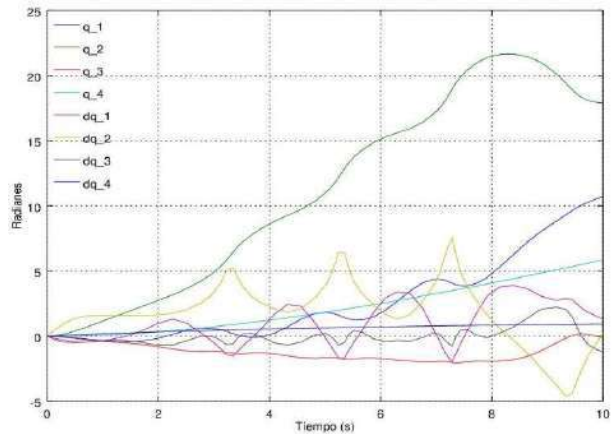

*Gráfico 2. Simulación en GNU-Octave del sistema no lineal: robot SCARA. Fuente: Elaboración propia.*

### b. Simulación del Péndulo Invertido Rotacional.

Utilizando el mismo procedimiento que se empleó en la simulación del robot SCARA, para la simulación del Péndulo Invertido Rotacional se utilizó el solucionador ode45, así como un tiempo de simulación de 5 segundos y condiciones inciales puestas a cero. Tanto para las simulaciones en Matlab$^{TM}$ como en GNU-Octave.

Los parámetros del péndulo invertido rotacional utilizados en el modelo de las ecuaciones diferenciales (8) y (9), son dados en la Tabla 2.

| Pará-metro | Valor | Pará-metro | Valor |
|---|---|---|---|
| $m_1$ | $0.2866\ kg$ | $L_0$ | $0.201\ m$ |
| $L_1$ | $0.30997\ m$ | $l_1$ | $0.15498\ m$ |
| $I_0$ | $0.0052$ | $I_1$ | $0.0023$ |

*Tabla 2: Parámetros del péndulo invertido rotacional. (Fuente: Elaboración propia).*

Los resultados de cada una de las simulaciones se visualizan a través de los Gráficos 3 y 4.

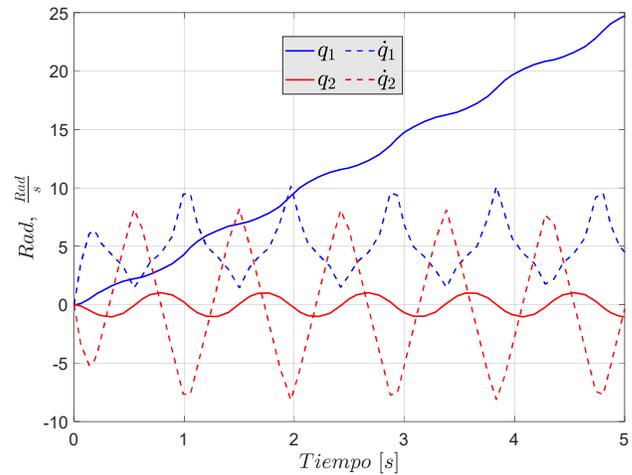

*Gráfico 3. Simulación en Matlab$^{TM}$ del sistema no lineal: péndulo invertido rotacional. Fuente: Elaboración propia.*

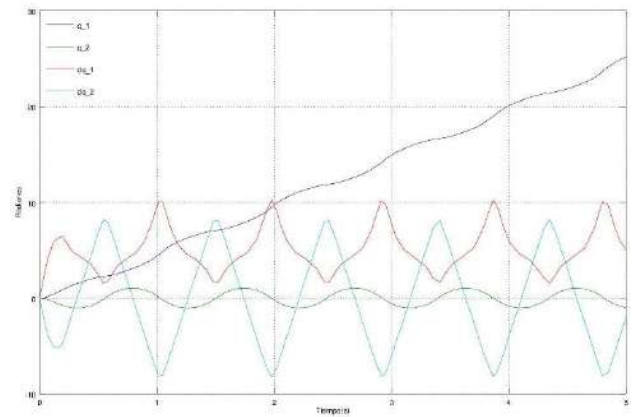

*Gráfico 4. Simulación en GNU-Octave del sistema no lineal: péndulo invertido rotacional. Fuente: Elaboración propia.*

### V. Conclusiones.

En este artículo se ha mostrado una alternativa para la enseñanza de la ingeniería cuando es necesario realizar simulaciones de los sistemas dinámicos no lineales para comprender el movimiento generado por la solución de sus ecuaciones diferenciales.

Sin embargo, para lograr un resultado eficiente, las simulaciones comúnmente se llevan a cabo en software especializado como Matlab$^{TM}$ que requiere de una licencia para su legal uso. La licencia por sí sola es un costo adicional para la institución educativa. Más aún, éste tipo de software necesita ser soportado en un equipo de cómputo con capacidades de hardware superiores a las computadoras de uso normal.

La alternativa mostrada consiste en usar GNU-Octave, el cual es un software libre que no requiere pagar el costo de una licencia para su legal funcionamiento.

Además, se ha demostrado que es posible utilizar una minicomputadora o tarjeta de desarrollo comercial, Raspberry Pi, como el dispositivo de cómputo sobre el cual está soportado el software GNU-Octave. Las comparaciones de las gráficas presentadas manifiestan, que la alternativa propuesta es viable para eficientar los recursos de la enseñanza de la ingeniería.